\documentstyle[11pt,epsf]{article}

\newfont{\subsub}{cmr6}

\newcounter{szk}

\begin{document}
\title{Relations between a typical scale and averages \\
in the breaking of fractal distribution
}
\author{
\footnote{e-mail address: ishikawa@kanazawa-gu.ac.jp} Atushi Ishikawa$^a$ and 
\footnote{e-mail address: tadao@nanao-c.ac.jp} Tadao Suzuki$^b$ \\
$^a$ Kanazawa Gakuin University, Kanazawa 920-1392, Japan\\
$^b$ Nanao Junior College, Nanao 926-8570, Japan
}
\date{}
\maketitle

\begin{abstract}
\indent

We study distributions which have both fractal and
non-fractal scale regions
by introducing a typical scale into a scale invariant system.
As one of models in which distributions follow power law in the large scale region
and deviate further from the power law in the smaller scale region,
we employ 2-dim quantum gravity modified by the $R^2$ term.
As examples of distributions in the real world which have similar property to this model,
we consider those of personal income in Japan over latest twenty fiscal years.
We find relations between the typical scale and several kinds of averages
in this model, and
observe that these relations are also valid in recent personal income distributions
in Japan with sufficient accuracy.
We show the existence of the fiscal years so called bubble term
in which the gap has arisen in power law,
by observing that the data are away from one of these relations.
We confirm, therefore, that the distribution of this model has close similarity 
to those of personal income.
In addition, we can estimate the value of Pareto index and
whether a big gap exists in power law 
by using only these relations.
As a result,
we point out that the typical scale is an useful concept 
different from average value
and that the distribution function derived in this model
is an effective tool to investigate these kinds of distributions. 
\end{abstract}

\begin{flushleft}
\end{flushleft}

\vspace{1cm}
\section{Introduction}
\label{sec-introduction}
\indent

Not only in nature but also in social phenomena,
various kinds of fractal structures 
are observed \cite{Mandelbrot}.
In a complete self-similar fractal system,
similar patterns are repeated in all scales of the system.
Even if we gaze a certain part of the system, therefore,
we cannot recognize the scale of the part. 
Fractal is the system without a typical scale
and the system is invariant under scale transformation.
This is equivalent to that
the distribution of the system follows power law. 

On the other hand,
many fractal structures appear in some restricted scales. 
In these distributions, the scale exists at which the fractal power law breaks.
The study for the breaking of fractal is as important as 
the study for fractal. 
Because most elements which constitute such distribution belong to a non-fractal region. 
There are various distributions which have both fractal and non-fractal scale regions
in phenomena which have no relation each other apparently.
If these distributions can be understood through some 
universality which does not depend on
the details of an individual system,
it may be possible that we can explain those in a unified way
by using an appropriate model.
For instance, L\'{e}vy's stable distribution \cite{Levy}
and Tsallis's $q$-Gaussian distribution \cite{Tsallis}
can explain the distribution which follows power law in the large scale region
and deviates from the power law in the other region.

In this paper, we take a following strategy.
It is mathematically simple to treat scale free theory.
One of the simplest methods to deal with the distribution which has
fractal and non-fractal scale regions systematically, therefore,
is to introduce a scale into the model, which is originally scale free,
by adding an interaction term with a scale:
\begin{eqnarray}
    S_{total}=S_{free}+S_{scale}~.
    \label{action total}
\end{eqnarray}
Here, $S_{free}$ is the scale free action, $S_{scale}$ is the interaction term with a scale
and $S_{total}$ is the total action.

There might be various candidates for $S_{free}$ and $S_{scale}$.
As one of kinds of the breaking of fractal,
there is the distribution which follows power law in the large scale region and
deviates further from the power law in the smaller scale region.
We employ 2-dim quantum gravity coupled with conformal matter field
as a model which can describe such distribution.
We take the standard scale invariant 2-dim action 
and the $R^2$ term with a scale
as $S_{free}$ and $S_{scale}$, respectively.
Here, the $R^2$ term is non-trivial and simplest interaction term with a scale.
Because the geometric characters of $S_{free}$ and $S_{scale}$
are known in 2-dim quantum gravity,
we can intuitively understand the behavior of fractal and 
the breaking of fractal. 

\begin{figure}[htb]
 \begin{minipage}{0.43\textwidth}
  \begin{center}
   \epsfysize=50mm
   \leavevmode
   \epsfbox{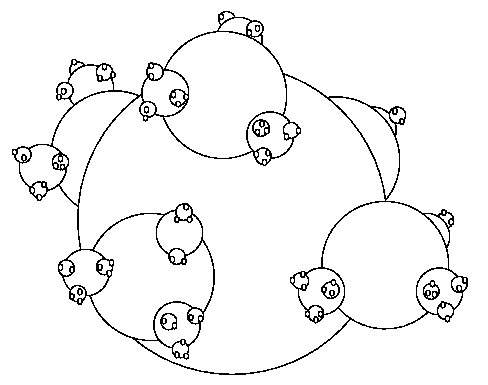}
  \end{center}
  \vspace{-5mm}
  \caption{A fractal 2-dim surface.}
  \label{fig:2-dim Random Surface}
 \end{minipage}
 \begin{minipage}{0.55\textwidth}
  \begin{center} 
   \epsfysize=50mm
   \leavevmode
   \epsfbox{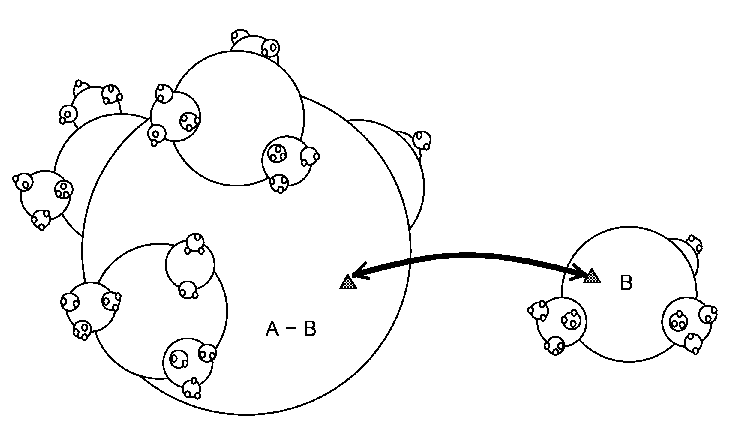}
   \caption{A 2-dim surface is divided into two MINBs.}
   \label{fig:Divided MINBs}
  \end{center}
 \end{minipage}
\end{figure}

In this model,
the scale free action $S_{free}$ leads
the locality that local structure of 2-dim surface does not 
affect the 2-dim structure away from that place. From this property,
it is well known that a typical 2-dim surface has self-similar structure
(Fig.~\ref{fig:2-dim Random Surface}) \cite{BIPZ, KAD}.
On the other hand,
the scale variant action $S_{scale}$ has the effect to
let 2-dim surface flat.
When this effect cannot be ignored, 
local structure of 2-dim surface does
affect the 2-dim structure away from that place.
The locality of 2-dim surface breaks down, therefore,
some kind of constraint is imposed on 2-dim surface.
As a result, 
the fractal structure in the small area region breaks down
and the typical scale with which fractal collapses is introduced into the system
\cite{KN, ITY}. 

In Ref.~\cite{AIST},
we examined whether the distribution functions obtained in this model 
could be applied to the distributions in nature or social phenomena. 
As examples of those which follow power law 
in the large scale region and deviate further from the power law
in the smaller scale region,
we considered two distributions of personal income \cite{Pareto, Gibrat, Badger}
and citation number of scientific papers \cite{Render}
respectively.
We showed that these distributions were fitted fairly well
by the distribution derived analytically in this model.
We also found that the values of the typical scales were comparable with
the average values of the distributions, 
and concluded that this model worked
consistently.
Because the quantity which has a scale in the whole distribution
does not exist except the average value.
This result might show the possibility that
we can explain these kinds of distributions universally
by using an appropriate model which has similar macroscopic characteristics,
not depending on the details of the model.
We cannot deny, however, the possibility that 
the distribution derived in this model are superficially similar to those of
personal income and there is no meaning beyond it.

In order to investigate this similarity of distributions more precisely,
in this paper,
we analyze the personal income distributions in Japan over recent twenty fiscal years.
We find relations between the typical scale and several kinds of averages
of the distribution in this model.
We observe that these relations are also valid with sufficient accuracy
in the latest personal income distributions in Japan over two decades years. 
We also show, in the intelligible form, 
the existence of the fiscal years in the bubble term,
the data of which are thought to be away from one of these relations
by the gap arisen in power law.
We insist, therefore, that this similarity between the distribution obtained in this model 
and the personal income distribution is very strict. 
In addition, even if we do not have detailed distributions in power law,
we can guess the value of Pareto index and
whether a big gap exists in power law 
by using only these relations.
As a result, 
we point out that 
the typical scale is an useful concept, which is different from average value,
and the distribution function derived in this model
is an effective tool to investigate the
distributions which have the similar property to this model.

\section{2-dim quantum $R^2$ gravity and dynamical triangulation}
\label{sec-2-dimensional}
\indent

As mentioned in Sec.~\ref{sec-introduction}, 
we take the standard 2-dim quantum gravity action
and the scale variant $R^2$ term 
as a scale free action $S_{free}$ and a scale dependent action $S_{scale}$,
respectively;
\begin{eqnarray}
    S_{free}(X^i;g)&=&\frac{1}{8 \pi} \int {\rm d^2}x \sqrt{g} g^{\mu \nu} 
    \partial_{\mu} X^i \partial_{\nu} X^i~,
    \label{action free}\\
    S_{scale}(X^i;g)&=&\frac{w}{64 \pi^2} \int {\rm d^2}x \sqrt{g}\, R^2~.
    \label{action scale}
\end{eqnarray}
Here $X^i(i=1,2,\cdots ,c)$ are conformal scalar matter fields coupled with gravity,
$g_{\mu \nu}(\mu, \nu=0, 1)$ is the metric of 2-dim surface,
$R$ is the scalar curvature and
$w$ is a coupling constant of length dimension $2$.
The partition function for fixed area $A$ of 2-dim surface 
is given by
\begin{eqnarray}
    Z(A)=\int \frac{{\cal D} g 
    {\cal D} X}{\rm vol(Diff)}
    {\rm e}^{-S_{total}(X^i;g)}~ \delta(\int d^2 x \sqrt{g}-A)~,
    \label{partition def}
\end{eqnarray}
where $\rm vol(Diff)$ is the volume of 2-dim
diffeomorphisms under which 
the action and the integration measure ${\cal D}g {\cal D}X$
are invariant. 
The asymptotic forms of the partition function
are evaluated as \cite{KN}
\begin{eqnarray}
Z(A)
&\sim& C_0~A^{\gamma_0-3}~\exp\biggl[
                    -\frac{w}{A}(1-h)^2 \biggr]
                            ~~~~~~~~~~{\rm for}~~A \ll w~,
\label{partition 0}
\\
&\sim& C_{\infty}~A^{\gamma_{\infty }-3}
                            ~~~~~~~~~~~~~~~~~~~~~~~~~~~~~~~~~
                            {\rm for}~~ A \gg w~,
\label{partition infty}
\end{eqnarray}
where $C_0$ and $C_\infty$ are proportional constants, and
$\gamma_0$ and $\gamma_{\infty}$ are constants determined by the central
charge $c$ and the number of handles of the 2-dim surface $h$,
\begin{eqnarray}
        \gamma_0(c, h) &=& \frac{(c-12)}{6}(1-h) + 2~,
\label{gamma_infty}\\
    \gamma_{\infty}(c, h) &=& \frac{c-25-\sqrt{(25-c)(1-c)}}{12}(1-h)+2~.
\label{gamma_0}
\end{eqnarray} Form
these expressions of the asymptotic forms (\ref{partition 0}) and (\ref{partition infty}),
we can observe that fractal power law is broken 
in the region $A \le w$.
If we take only the standard scale free 2-dim action (\ref{action free})
without the scale variant $R^2$ term (\ref{action scale}),
the theory does not contain the scale $w$, and the partition function
follows the fractal power law (\ref{partition infty}) in all scale regions
\cite{BIPZ}.

In 2-dim quantum gravity, the method is established 
to observe this phenomenon in numerical simulation.
It is known as MINBU analysis \cite{JM}
in dynamical triangulation (DT) \cite{KAD}.
In DT, 2-dim surface is discretized using small equilateral triangles,
where each triangle has the same size.
The evaluation of the partition function is performed 
by replacing the path integral over the metric with the sum over 
possible triangulations of 2-dim surface. From various evidences,
DT is thought to be equivalent to the continuum theory of 2-dim gravity
if there are sufficient triangles \cite{Takusan}.

A MINBU (minimum-neck baby universe) 
is defined as a simply connected area region of 2-dim surface
whose neck is composed of three sides of triangles,
where the neck is closed and non-self intersecting.
In MINBU analysis
\footnote{Originally, the MINBU analysis is invented to measure
$\gamma_{\infty}$ in the numerical simulation.}, 
by dividing a closed 2-dim surface
into two MINBUs at a minimum neck
(Fig.~\ref{fig:Divided MINBs})
the statistical average number of
finding a MINBU of area $B$ on a closed surface of area $A$,
$n_A(B)$, can be expressed as
\begin{eqnarray}
 n_A(B)\sim \frac{B(A-B) Z(B)  Z(A-B)}{Z(A)}~.
\label{MINBU def}
\end{eqnarray}
Here we set the area of a triangle $a^2=1$ for simplicity.
Using the asymptotic forms of the partition functions 
(\ref{partition 0}) and (\ref{partition infty})
in the continuum theory,
we obtain the asymptotic expressions of $n_A(B)$
\begin{eqnarray}
 n_A(B)
&\sim& C_0~A~ 
      B^{\gamma _0 -2}
      \exp\biggl[ -\frac{w}{B}(1-h)^2 \biggr]
\qquad\qquad
      {\rm for }~~ 1 \ll B \ll w~,
 \label{MINBU Weibull}\\
&\sim& C_{\infty}~A~  \biggl[(1-\frac{B}{A})B \biggl]^{\gamma _\infty -2}
\qquad\qquad\qquad~~
      {\rm for }~~w \ll B < A/2~.
 \label{MINBU Pareto}
\end{eqnarray}
These MINBU distributions (\ref{MINBU Weibull}) and (\ref{MINBU Pareto})
keep same properties which partition functions 
(\ref{partition 0}) and (\ref{partition infty}) possess.
As for the case $w \ll B$,
the asymptotic form
(\ref{MINBU Pareto})
follows power law.
In this range, even if the model contains the scale $w$, 
at an area scale much larger than $w$, the surfaces are 
fractal. 
On the other hand, as for the case $B \ll w$,
the asymptotic form
(\ref{MINBU Weibull}) is highly suppressed by the
exponential factor $\exp\bigl[ -\frac{w}{B}(1-h)^2 \bigr]$.
In this range,  
at an area scale much smaller than $w$, the surfaces
are affected by the typical length scale $w$.
The distribution of smaller MINBUs are, therefore, is apart further from power law.
In this paper, we call the distribution (\ref{MINBU Weibull}) and the scale $w$
as Weibull distribution and Weibull scale, respectively.

In DT, the $R^2$ term
(\ref{action scale}) is expressed by
\begin{eqnarray}
    \int d^2 x \sqrt{g} R^2  
    \cong \frac{4 \pi^2}{3}  \sum_i 
      \frac{(6-q_i)^2}{q_i} ~.
\label{discretized R2}
\end{eqnarray}
Here $q_i$ is the number of triangles sharing the vertex $i$. From
Eq.~(\ref{discretized R2}), we can recognize that the $R^2$ term
has the effect 
to make 2-dim surface flat ($q_i=6$) and suppress
the possibility that
the small MINBUs exist.
This effect is parameterized by the coefficient of the $R^2$ term.
The method to measure MINBU distributions is well known in DT.
We can observe the breaking of the fractal by the typical scale
in the numerical way. 

\section{Numerical analysis of DT}
\label{sec-numerical}
\indent

\begin{figure}[htb]
 \centerline{\epsfxsize=0.78\textwidth\epsfbox{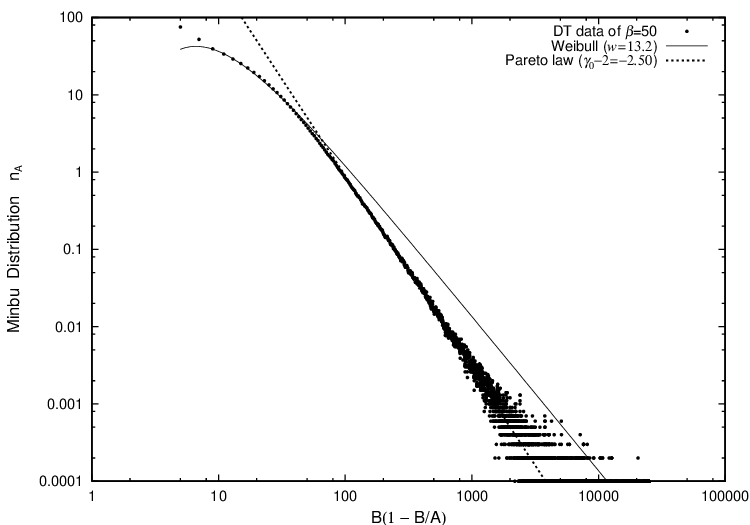}}
 \caption{The simulation data and the fitting of $\beta_L=50$.}
 \label{fig:beta=50}
\end{figure}

\begin{figure}[htb]
 \centerline{\epsfxsize=0.78\textwidth\epsfbox{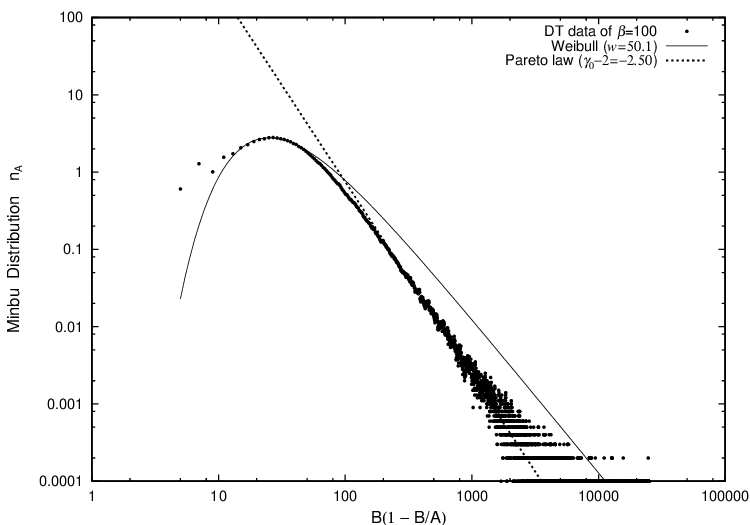}}
 \caption{The simulation data and the fitting of $\beta_L=100$.}
 \label{fig:beta=100}
\end{figure}

The asymptotic forms of MINBU distributions 
(\ref{MINBU Weibull}) and (\ref{MINBU Pareto}) are
analytically obtained in 
Sec.~\ref{sec-2-dimensional}.
They can be also confirmed in the simulation of DT
for the simple case that 2-dim surface is sphere ($h=0$)
and there is no matter field on it ($c=0$) \cite{ITY}. 
For example, two simulation results are represented  
in Figs.~\ref{fig:beta=50} and \ref{fig:beta=100},
where the total number of triangles is 100,000.
We plot MINBU distributions, $n_A(B)$ versus $(1-B/A)B$ 
with a log-log
scale for $\beta_L = 50$ and $100$, which are coefficients of 
the discretized $R^2$ term (\ref{discretized R2}).
Here, we can replace $B$ with $(1-B/A)B$ in Eq.~(\ref{MINBU Weibull}),
because 
$B \ll w \ll A$. 
These MINBU distributions can be well explained 
by the asymptotic formulae (\ref{MINBU Weibull}) and (\ref{MINBU Pareto})
with $\gamma_0=0$ and $\gamma_\infty=-1/2$
($h=c=0$).
The data fittings for $\beta_L=50$ and $100$
are also expressed in Figs.~\ref{fig:beta=50} and \ref{fig:beta=100},
respectively
\footnote{
Several data points for small MINBUs are
apart from the Weibull distribution (\ref{MINBU Weibull}).
We consider that it is the finite lattice effect.
}. From Figs.~\ref{fig:beta=50} and \ref{fig:beta=100},
we observe that the asymptotic forms (\ref{MINBU Weibull}) and (\ref{MINBU Pareto})
can be applicable to almost all regions of MINBU distributions.
These asymptotic representations, therefore, 
might be used as the approximations of MINBU distributions.
In next Sec.~\ref{sec-approximation}, we examine this possibility.

\section{Approximation of MINBU distributions}
\label{sec-approximation}
\indent

In this section, first we postulate the approximation forms of MINBU distributions,
and derive relations between Weibull scale $w$ and several kinds of averages
by using them.
Next, we examine whether these relations are valid 
in DT simulations. 
After here,
we set $h=0$ for simplicity. 

From the asymptotic formulae (\ref{MINBU Weibull}) and (\ref{MINBU Pareto}),
we assume that the approximations of MINBU distributions are 
\begin{eqnarray}
   n_W(x)&\sim& N_W~x^{\gamma_0-2}~
   \exp\biggl[ -\frac{w}{x}
   \biggr]
   ~~~~~~   {\rm for}~~~~ 0 \le  x \le w ,
   \label{Weibull Aproximation}\\
   n_P(x)&\sim&  N_P~x^{\gamma_{\infty}-2}   ~~~~~~~~~~~~~~~~~~~~
   {\rm for}~~~~ w \le x < \infty~,
   \label{Pareto Aproximation}
\end{eqnarray}
where $x=(1-B/A)B$ and 
$N_W$ and $N_P$ are normalization constants.
Using these forms,
we calculate an
average of the distribution in power law region
$\left\langle(x_m \le)~x~(\le x_M)\right\rangle_P$, 
one in Weibull range 
$\left\langle x~(\le x_M)\right\rangle_W$
and one in all regions 
$\left\langle (x_m \le)~x \right\rangle_{W+P}$
as follows, respectively:
\begin{eqnarray}
   \left\langle(x_m \le)~x~(\le x_M)\right\rangle_P&\equiv& 
    \frac{\int_{x_m}^{x_M} dx~n_P(x)~x}{\int_{x_m}^{x_M} dx~n_P(x)}\nonumber\\
        &\sim& \frac{1-\gamma_{\infty}}{-\gamma_{\infty}}
            \left\{1-\left(\frac{x_m}{x_M}\right)^{-\gamma_{\infty}} \right\}x_m
    ~~~{\rm for}~~~\frac{x_m}{x_M} \ll 1~,\\
   \left\langle x~(\le x_M)\right\rangle_W&\equiv&
   \frac{\int_{0}^{x_M} dx~n_W(x)~x}{\int_{0}^{x_M} dx~n_W(x)}
        =\frac{\Gamma(-\gamma_0, \frac{w}{x_M})}{\Gamma(1-\gamma_0, \frac{w}{x_M})}w,
        \label{average1}\\
   \left\langle (x_m \le)~x \right\rangle_{ W+P}&\equiv&
    \frac{\int_{x_m}^{w} dx~n_W(x)~x + \int_{w}^{\infty} dx~n_P(x)~x}
        {\int_{x_m}^{w} dx~n_W(x)+\int_{w}^{\infty} dx~n_P(x)}
        \nonumber \\
   &=& \frac{e~\Gamma(-\gamma_0,1)+\frac{1}{-\gamma_0}}
        {e~\Gamma(1-\gamma_0,1)+\frac{1}{1-\gamma_0}} w
            +\frac{1-\gamma_{\infty}}{-\gamma_{\infty}}x_m~\label{average2}.
\end{eqnarray}
Here, $\Gamma(z,p)$ is incomplete gamma function defined by
$\Gamma(z,p)=\int^{\infty}_{p} dt~e^{-t}~t^{z-1}$.
We used the continuous condition $n_W(w)=n_P(w)$ to calculate Eq.~(\ref{average2}).

In the simulation of DT
for the simple case that 2-dim surface is sphere
and there is no matter field on it ($\gamma_0=0$, $\gamma_\infty=-1/2$),
these relations 
can be represented as
\begin{eqnarray}
   \left\langle (x_m \le)~x \right\rangle_{ W+P}
   &\sim&\frac{3}{5} \left(
        e \Gamma(0,1)+2 \right)w+3x_m
        ~~~~~~~~~~~~~~~~~~~~~{\rm for}~~~\frac{w}{x_M} \ll 1
        \nonumber \\
        &\sim& 1.56~w +3~x_m~,\label{approximation-3}\\
   \left\langle x~(\le x_M)\right\rangle_W
        &\sim& e~\Gamma(0,1)~w - \left\{ 1-e~\Gamma(0,1) \right\}(w-x_M)
    ~~~{\rm for}~~~1-\frac{x_M}{w} \ll 1
   \nonumber \\ 
        &\sim& 0.596~w - 0.404(w-x_M)\label{approximation-1}\\
        &=&0.192~w + 0.404~x_M~.\label{approximation-2}
\end{eqnarray}

\begin{figure}[htb]
 \centerline{\epsfxsize=0.78\textwidth\epsfbox{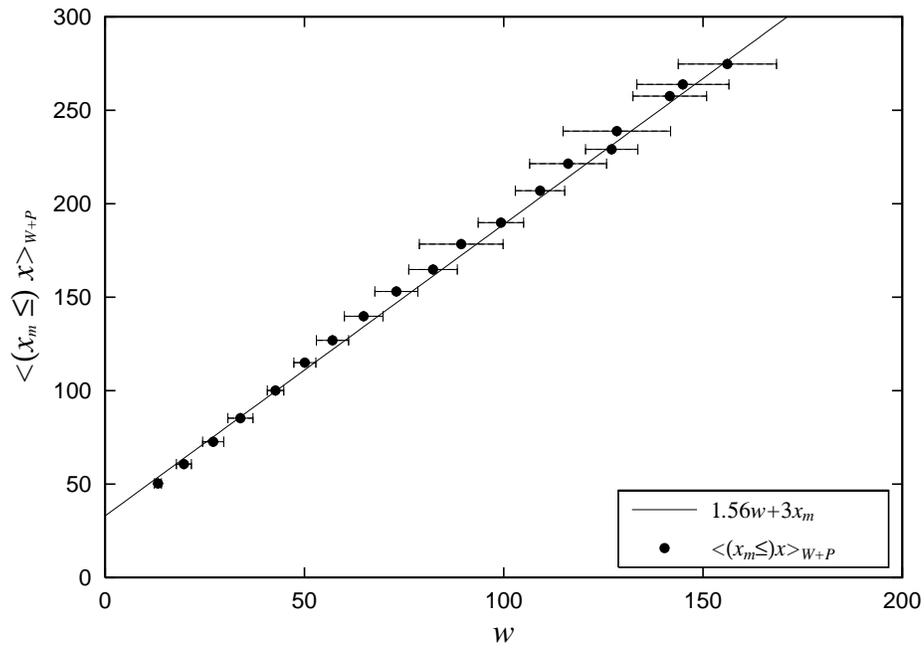}}
 \caption{The relation between Weibull scales $w$ and 
 average values of the whole distribution 
 $\left\langle (x_m \le)~x \right\rangle_{W+P}$ in DT simulations.}
 \label{Relation-1 in Simulation}
\end{figure}

\begin{figure}[htb]
 \centerline{\epsfxsize=0.78\textwidth\epsfbox{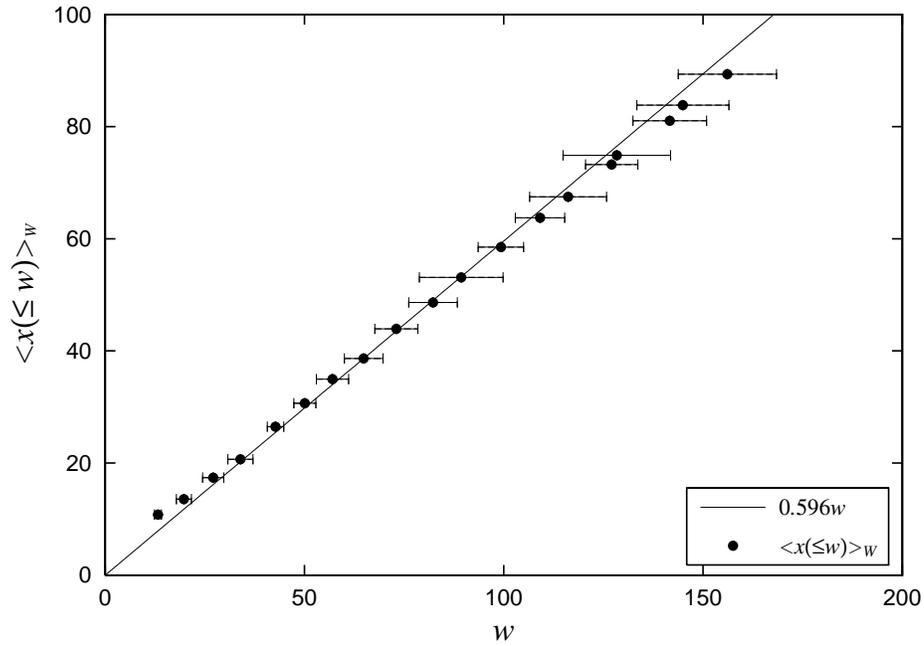}}
 \caption{The relation between Weibull scales $w$ and 
 average values below the Weibull scale $\left\langle x~(\le w) \right\rangle_W$ in DT simulations.}
 \label{Relation-2 in Simulation}
\end{figure}

\begin{figure}[htb]
 \centerline{\epsfxsize=0.78\textwidth\epsfbox{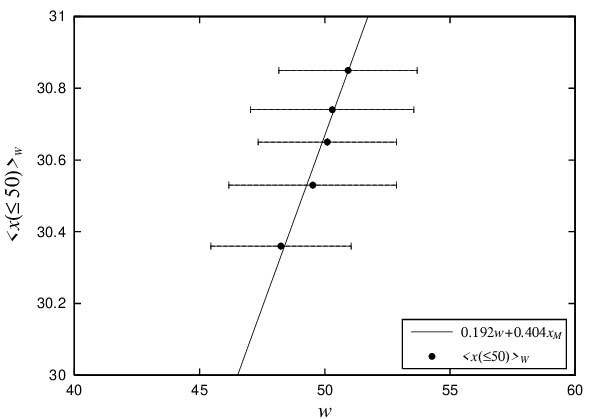}}
 \caption{The relation between Weibull scales $w$ and
 average values below MINBU area $50$ $\left\langle x~(\le 50) \right\rangle_W$ in DT simulations.}
 \label{Relation-3 in Simulation}
\end{figure}

Analyzing results of DT simulations,
we confirm that these relations,
which are obtained by using the approximations 
(\ref{Weibull Aproximation}) and (\ref{Pareto Aproximation}), 
are well accurate.
First, 
we express the simulation results and the theoretical line
with respect to Eq.~(\ref{approximation-3}) 
in Fig.~\ref{Relation-1 in Simulation}.
The horizontal axis indicates Weibull scale
and the vertical axis indicates the average.
Second, 
Weibull scale $w$ correlates with $\left\langle x~(\le w)\right\rangle_W$
for the case $x_M=w$ in Eq.~(\ref{approximation-1}).
We represent the simulation results and the theoretical line
with respect to this correlation
$\left\langle x~(\le w)\right\rangle_W \sim 0.596~w$
in Fig.~\ref{Relation-2 in Simulation}.
We also use 100,000 triangles,
and plot simulation data for $\beta_L = 50$, $60$, $70$, $\cdots$, $230$
in these two kinds of simulations.
Third, the simulation results and the theoretical line
with respect to Eq.~(\ref{approximation-2})
are expressed in Fig.~\ref{Relation-3 in Simulation}. 
We fix $x_M=50$ and use the values of $w$ near the $x_M$
for $\beta_L = 98$, $99$, $100$, $101$, $102$.

In every cases, the theoretical lines with respect to
Eq.~(\ref{approximation-3}), (\ref{approximation-1}) and (\ref{approximation-2})
explain the simulation data with sufficient accuracy.
As a result,
we recognize Eqs.~(\ref{Weibull Aproximation}) and (\ref{Pareto Aproximation})
as good approximations to calculate the relations
between these averages and Weibull scale.

\section{Personal Income Distributions}
\label{sec-personal}
\indent

In this section, we point out that the relations in this model are also valid
in the personal income distributions 
interested in Econophysics \cite{MS}. 
We analyze the fiscal years 1980 - 2000. 

In Japan, persons who paid the income tax more than 10 million yen
are announced publicly as "high income taxpayers" every year. 
It is difficult to get the complete data,
but we can procure some data from some company.
It is possible to estimate the personal income distribution of "high income taxpayers"
by using the method in Ref.~\cite{ASNOTT}.
"The rough distributions of less than 50 million yen income earners" 
are also released by the National Tax Administration Agency in Japan. 
These are comparatively easy to get and we can download latest data 
from web page \cite{Tax URL}.
We can obtain the distribution of all income earners
combining these two kinds of data.

\begin{figure}[htb]
 \centerline{\epsfxsize=0.78\textwidth\epsfbox{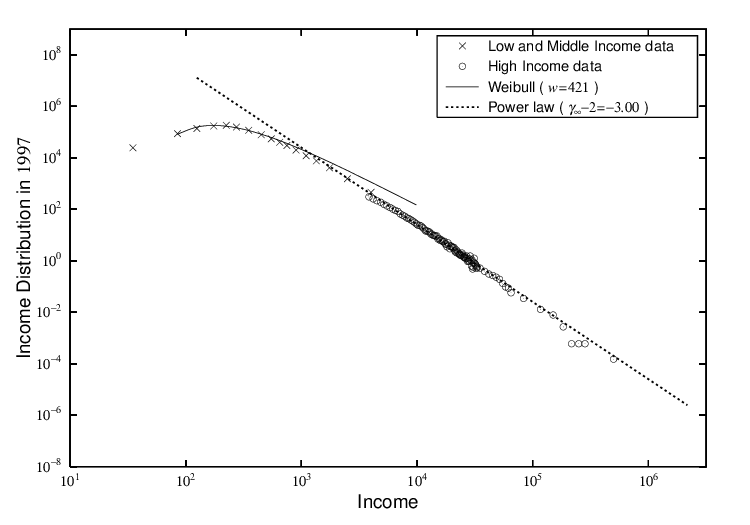}}
 \caption{The personal income distribution and the data fitting in 1997 Japan.}
 \label{Personal Income-1}
\end{figure}

\begin{figure}[htb]
 \centerline{\epsfxsize=0.78\textwidth\epsfbox{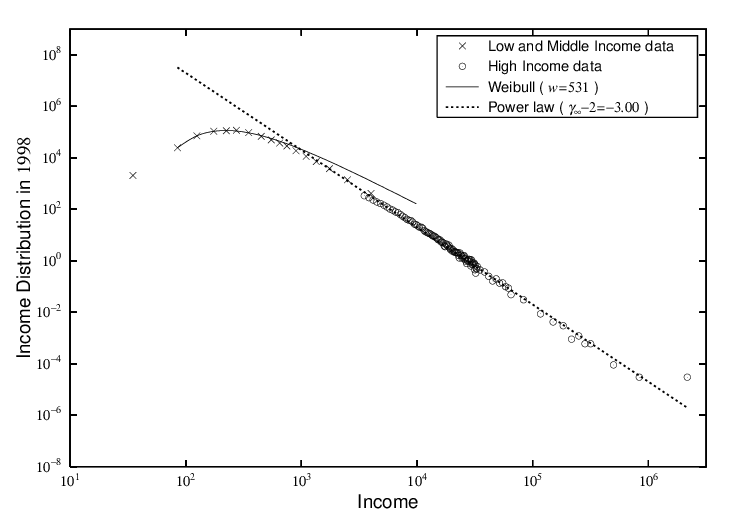}}
 \caption{The personal income distribution and the data fitting in 1998 Japan.}
 \label{Personal Income-2}
\end{figure}

In Ref.~\cite{AIST}, we examined whether the distributions derived in this model
could be applicable to the 
personal income distributions in 1997 and 1998 Japan
in which we procured "high income taxpayers"
data from Ref.~\cite{Chouja}. 
We observed that $\gamma_\infty-2 = -3$ in the power law region (\ref{MINBU Pareto}),
then we decided that $\gamma_0 - 2 = -7/3$ in Weibull region (\ref{MINBU Weibull}).
These parameters are realized by setting that $c=-2$ and $h=0$.
We showed that the personal income distributions were also fitted fairly well
by the distributions (\ref{MINBU Weibull}) and (\ref{MINBU Pareto}),
and the typical scales could be read consistently
(Figs.~\ref{Personal Income-1} and \ref{Personal Income-2}).
In these Figs, 
the horizontal axis indicates the income $x$ in units of 10 thousand yen
and the vertical axis indicates the number density of persons per a period of
100 thousand yen.

\begin{figure}[htb]
 \centerline{\epsfxsize=0.78\textwidth\epsfbox{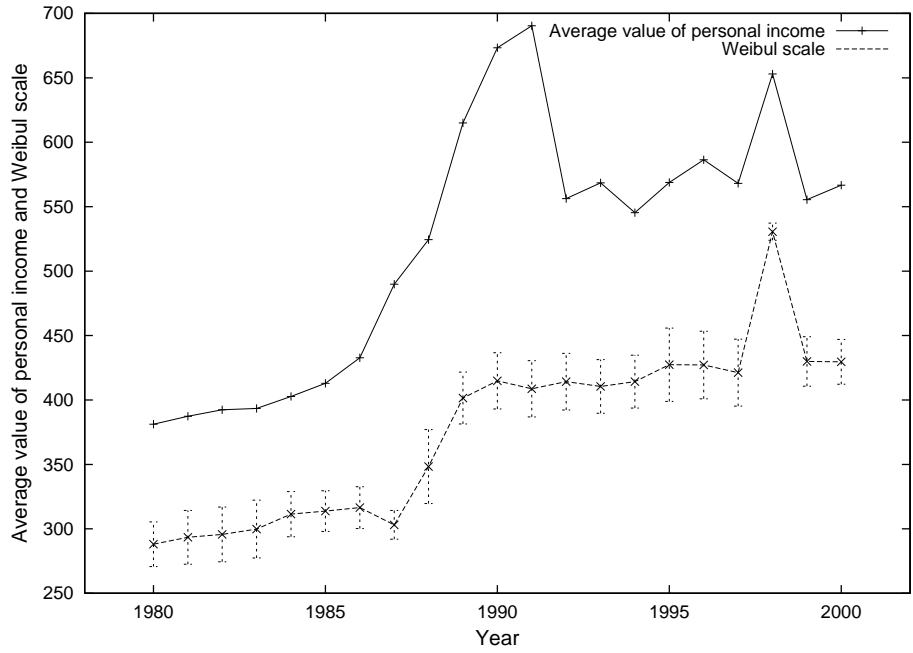}}
 \caption{Weibull scales and average values of the personal income in 1980 - 2000 Japan.}
 \label{Average and Weibull Scale}
\end{figure}

Recently, in Ref.~\cite{FSAKA}, it is reported that the value
$\gamma_\infty-2$ is almost $-3$ (Pareto Index $\alpha=2$)
in 1987 - 2000 Japan.
Combining this report and "the rough distribution" 
released by the National Tax Administration Agency,
we can calculate Weibull scales in these fiscal years
in a same mannar in 1997 or 1998 Japan.
We can also estimate income average values by using the data with respect to 
the sum of income and the total number of income earners
which the National Tax Administration Agency has also announced
(Fig.~\ref{Average and Weibull Scale}). 
We examine whether the relations 
between Weibull scale and several kinds of averages,
which is verified in Sec.~\ref{sec-approximation},
are valid in these data of the personal income distributions in recent Japan.
In order to consider fiscal years before, in and after bubble term in Japan continuously,
we analyze the fiscal years 1980 - 2000 assuming that
the value $\gamma_\infty-2$ is almost $-3$ in 1980 - 1986
likely in 1987 - 2000.

Using the approximate distribution functions 
(\ref{Weibull Aproximation}) and (\ref{Pareto Aproximation}), from Eqs.~(\ref{average1}) and (\ref{average2})
for the case that $\gamma_0=-1/3$ and $\gamma_\infty=-1$
we obtain the average
in all regions 
$\left\langle (x_m \le)~x \right\rangle_{W+P}$
and one in Weibull range 
$\left\langle x~(\le x_M)\right\rangle_W$
as follows, respectively:
\begin{eqnarray}
   \left\langle (x_m \le)~x\right\rangle_{W+P}
        &\sim& 0.980~w +2~x_m~,
        \label{personal2}\\
   \left\langle x~(\le x_M)\right\rangle_W
        &=&\frac{\Gamma(\frac{1}{3}, \frac{w}{x_M})}{\Gamma(\frac{4}{3}, \frac{w}{x_M})}~w \nonumber \\
        &\sim& 0.566~w - 0.353(w-x_M) ~~~{\rm for}~~~1-\frac{x_M}{w} \ll 1 \label{personal0}\\
        &=&0.213~w + 0.353~x_M~.
           \label{personal}
\end{eqnarray}

\begin{figure}[htb]
 \centerline{\epsfxsize=0.78\textwidth\epsfbox{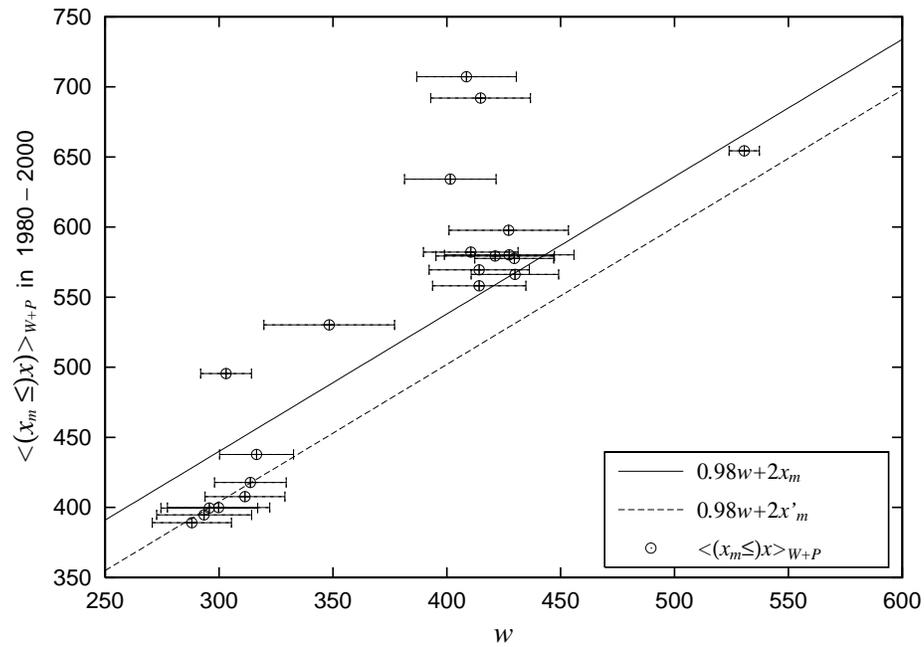}}
 \caption{The relation between Weibull scales $w$ and average values of the whole personal income
 $\left\langle (x_m \le)~x \right\rangle_{W+P}$ in 1980 - 2000 Japan.}
 \label{Average and Weibull Scale-1}
\end{figure}

We show average values in whole distribution and Weibull scales,
and also express the theoretical lines (\ref{personal2}) in 
Fig.~\ref{Average and Weibull Scale-1}.
Because classification of data in 1980 - 1988 is different from
one in 1989 - 2000,
there are two theoretical lines corresponding to two different values of $x_m$.
These values are represented in units of 10 thousand yen.
It turns out that the data of 1980 - 1986 and 1992 - 2000 fiscal years 
agree with each theoretical line respectively,
and that those of 1987 - 1991 fiscal years disagree with either.
In Ref.~\cite{FSAKA}Cit is also reported that
the gap, which cannot be disregarded, in power law
has arisen in the 1987 - 1991 fiscal years, so called bubble term.
Error bars in Fig.~1 in Ref.~\cite{FSAKA}
indicate the gaps.
Our model presupposes the power law in the large scale region.
The distribution function in this model, therefore,
cannot be applicable to the distributions in this term strictly and 
the relation between average values in whole distribution and Weibull scales 
(\ref{personal2}) is not maintained.
Moreover, from the result that this relation keeps in 1980 - 1986,
we guess that there are not big gaps in the power law
and the value $\gamma_\infty-2$ is almost $-3$ in this term.

\begin{figure}[htb]
 \centerline{\epsfxsize=0.78\textwidth\epsfbox{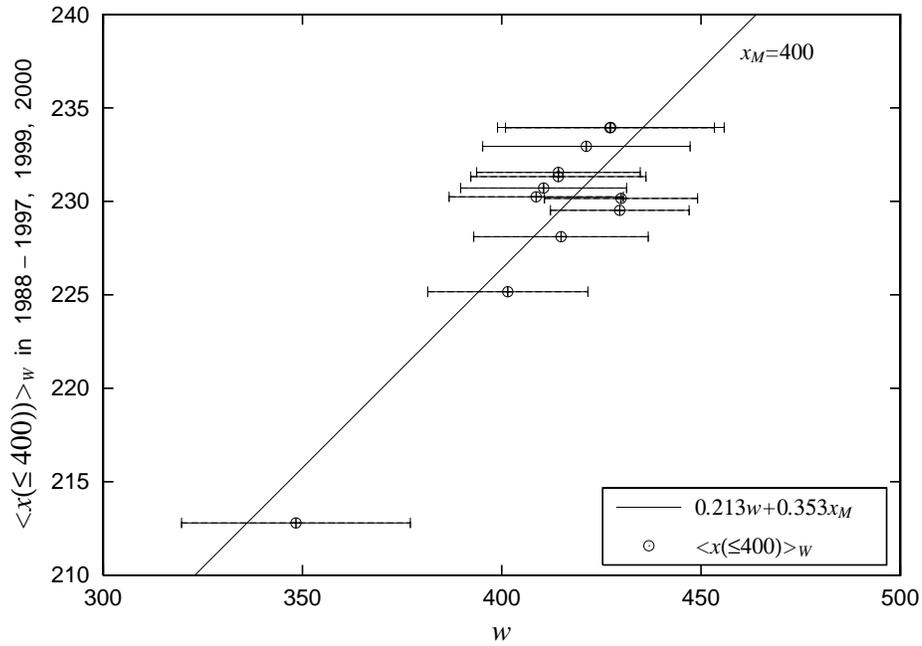}}
 \caption{The relation between Weibull scales $w$ and average values of the personal income
 below $4$ million yen $\left\langle x~(\le 400) \right\rangle_W$ in 1988 - 1997, 1999, 2000 Japan.}
 \label{Average and Weibull Scale-2}
\end{figure}

\begin{figure}[htb]
 \centerline{\epsfxsize=0.78\textwidth\epsfbox{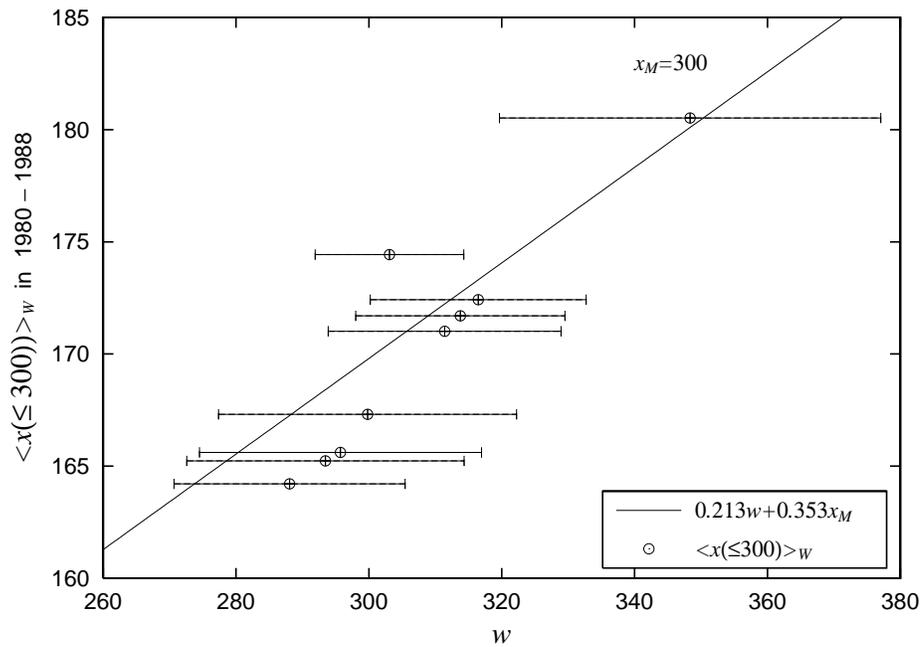}}
 \caption{The relation between Weibull scales $w$ and average values of the personal income
 below $3$ million yen $\left\langle x~(\le 300) \right\rangle_W$ in 1980 - 1988 Japan.}
 \label{Average and Weibull Scale-3}
\end{figure}

On the other hand, we cannot examine the relation 
$\left\langle x~(\le w)\right\rangle_W \sim 0.566~w$
in Eq.~(\ref{personal0}) for the case $x_M=w$,
because we have no data in which the value $x_M$ can be finely moved
according to the change of the value of $w$.
It is possible, however, to examine the relation (\ref{personal})
between Weibull scale $w$ and $\left\langle x~(\le x_M)\right\rangle_W$
by fixing the value of $x_M$.
In Fig.~\ref{Average and Weibull Scale-2},
the relation (\ref{personal}) for the case $x_M=400$ is confirmed
in 1988 - 1997, 1999, 2000 fiscal years
where Weibull scales are almost 4 million yen.
In Fig.~\ref{Average and Weibull Scale-3},
the relation (\ref{personal}) for the case $x_M=300$ is also confirmed
in 1980 - 1988 fiscal years
where Weibull scales are almost 3 million yen.
The data in 1987 - 1991 fiscal years,
where the gap has arisen in power law,
are not away from this relation (\ref{personal}).
It can be understood that 
the distribution in power law does not have great influence 
on the average in Weibull region.

We find that the relation (\ref{personal2}) 
is also valid 
in the distributions of personal income.
This means that 
the position of the typical scale in the whole distribution in our model
is very similar to the position 
in the personal income distributions. 
We also find that the relation (\ref{personal}) 
keeps in the personal income distributions.
This means that the two kinds of distributions in the region where
variables are smaller than the typical scale is also very alike. From 
these observations,
we conclude that
there is high similarity between the distribution derived in our model
and those of personal income in latest Japan,
and that our distribution function can be applicable to the personal income distribution.

\section{Summary and discussion}
\label{sec-summary}
\indent

We studied the breaking of fractal distribution
by introducing a typical scale into a scale invariant system
in which the distribution is fractal.
We employed 2-dim quantum gravity modified by the $R^2$ term
as the model which can describe the distribution
which follows power law in the large scale region
and deviates further from the power law in the smaller scale region.
We concentrated ourselves on considering the personal income distribution
which has similar property to this model in the real world.
In Ref.~\cite{AIST}, we only pointed out that
the distribution derived in this model can fit those of personal income
in 1997 and 1998 Japan.
In this paper, we found relations between the typical scale and several kinds of 
averages of the distribution in this model.
We observe that these relations are also valid with sufficient accuracy
in the personal income distributions in 1980 - 2000 Japan.
We also showed that the data in 1987 - 1991 fiscal years are apart from one of these relations.
This reason can be thought to be that gaps have arisen in the power law in these years.
This phenomenon is reported in Ref.~\cite{FSAKA}.
We also estimated that there are not big gaps in the power law in 1980 - 1986 fiscal years
and the values of Pareto index in these years are almost $2$.
This is concluded by the agreement of this relation with data in these years.
In this analysis, we do not need the details of the distribution of high income earners.
As a result, we consider that the typical scale is an useful concept
and that the distribution function in this model is an effective tool 
to investigate the distributions which have the similar property to this model.

We do not understand that this coincidence of the distributions
can be thought to be an universal property.
In other words,
we cannot refer to any physical connection between 2-dim $R^2$ gravity
and personal income.
The following speculations, however, might give some insight.
Fractal is observed in the system described by only independent variables.
This is realized in 2-dim gravity.
This situation is also expected to be realized in free competition
which leads the distribution of high income earners.
On the other hand, the breaking of fractal is observed in the system
with some kind of constraint.
This is realized in 2-dim gravity with the $R^2$ term
in which this constraint is induced by the interaction with 2-dim surface.
We expect that there is also some kind of interaction in middle income earners
\cite{SFA}.
These way of thinking might be clear 
by investigating the company's income distributions
which have more informations than those of personal income.


\section*{Acknowledgements}
\indent

The authors would like to express our gratitude to 
Professor.~H. Aoyama and Dr.~Y. Fujiwara for 
significant explanations about their work \cite{FSAKA}.
We are also grateful to the Yukawa Institute for Theoretical 
Physics at Kyoto University,
where this work was initiated during the YITP-W-03-03 on
"Econophysics - Physics-based approach to Economic and
Social phenomena -".
Thanks are also due to Dr.~M. Tomoyose and Professor.~T. Maeda
for useful comments and discussions.



\newpage


\end{document}